# Using HCI to Tackle Race and Gender Bias in ADHD Diagnosis

**Naba Rizvi**
University of Toledo
Toledo, OH 43606
nabah.rizvi@rockets.utoledo.edu

**Khalil Mrini**
University of California, San Diego
La Jolla, CA 92093
khalil@ucsd.edu

## Abstract
Attention Deficit Hyperactivity Disorder (ADHD) is a behavioral disorder that impacts an individual's education, relationships, career, and ability to acquire fair and just police interrogations. Yet, traditional methods used to diagnose ADHD in children and adults are known to have racial and gender bias. In recent years, diagnostic technology has been studied by both HCI and ML researchers. However, these studies fail to take into consideration racial and gender stereotypes that may impact the accuracy of their results. We highlight the importance of taking race and gender into consideration when creating diagnostic technology for ADHD and provide HCI researchers with suggestions for future studies.

## Author Keywords
ADHD; behavioral disorder; mental disorder; bias; psychology; racial bias; gender bias; healthcare bias.

## Identifying Bias in ADHD Diagnosis

Healthcare researchers have emphasized the importance of developing universal diagnostic tools to combat bias in ADHD diagnosis [15].

**Race:** Children of color are up to 69% less likely than white children to receive an ADHD diagnosis due to racial stereotypes or lack of resources [2][3][15][18]. However, tools proposed by computer scientists have not studied the impact race may have on their results [18][19][20][21].

**Gender**: Women are more likely to have inattentive ADHD and compensate for their disorder through adapting prosocial behaviors [15][22]. Thus, women diagnosed with ADHD may have more severe symptoms. Yet, none of the studies stated how this may have impacted their results by introducing selection bias [19][20][21][22].

## Introduction

According to the Center for Disease Control & Prevention, Attention Deficit Hyperactivity Disorder (ADHD) is one of the most prevalent mental health disorders affecting children. In the United States, five million children aged 3–17 were diagnosed with it in 2010 and boys were twice as likely to get diagnosed than girls [1]. ADHD is a behavioral disorder that impacts an individual's academic performance, relationships, self-esteem, and quality of life. Yet, compared to white children, black, Hispanic, or children from other races are respectively 69%, 50%, and 46% less likely to receive a diagnosis for ADHD [2]. Researchers have applied the critical race theory to study the racial differences in diagnosing Oppositional Defiant Disorder (ODD) and ADHD. They uncovered that despite a similar prevalence of both disorders among racial groups, the myth of the "dangerous black child" caused black children to be over-diagnosed with ODD instead of ADHD [3]. Additionally, prior research estimates that around half to three-quarters of women with ADHD are undiagnosed and girls receive a diagnosis 5 years later than their male counterparts on average [4]. Technology provides quantitative methods of evaluating psychiatric disorders which may eliminate racial and gender bias thereby making a timely ADHD diagnosis affordable and accessible to people who have been historically underdiagnosed.

## Why It Matters

Underdiagnosis of ADHD can have serious consequences on the lives of people of color and women. The disorder is linked to delinquent behavior in youth and adults. People with ADHD comprise an estimated 70% of juvenile offenders and 40% of adult prisoners overall [5]. Additionally, the underdiagnosis of ADHD in black boys may contribute to their over-representation in the "school-to-prison pipeline" [6]. One study discovered that people with ADHD are psychologically vulnerable in police questioning and may become victims of peer pressure and false confessions because of their poor concentration, restlessness, and impulsiveness [7]. The same study also expanded how ADHD impacts an individual's education. Children with the disorder may be placed into curriculums for "low-achievers" which impacts their self-esteem and causes them to live up to this negative expectation with negative social and academic behavior. Students with ADHD face an increased risk of suspension, expulsion, or dropping out [7]. Other impacts of the disorder on individuals include: facing increased disciplinary action in schools [8], an increased risk of smoking in female adolescents [9], sleep disorders in both children and adults [10], poor interpersonal skills that lead to fewer friendships and more remarriages [11], and a greater rate of changing jobs due to dismissals or their own will [12]. One study reviews the use of various intervention tools in classrooms and society [7]. The study highlights the importance of early diagnosis and intervention in supporting children and youth with ADHD and improving their quality of life.

## Traditional Diagnostic Methodologies

The Diagnostic and Statistical Manual of Mental Disorders (DSM) is used by healthcare professionals in the United States as the authoritative guide on mental health disorders [13]. Previously, the manual was used to create questionnaires and criteria used in diagnosis of ADHD. This method of diagnosis has its limitations as it largely relies on the observations and opinions of healthcare professionals, educators, and family members of patients which can be biased [14]. For example, parents of black or Latino children may not recognize the symptoms of the disorder or find resources that are culturally relevant which causes healthcare providers to not screen their children for ADHD [15]. Additionally, the symptoms of the disorder can be situational and may vary based on gender. For example, boys are more likely than girls to display symptoms of hyperactivity, whereas girls are more likely to be inattentive [16]. Furthermore, diagnosis can be even more difficult for adults. Overachieving college students may have "positive illusory bias" which can cause them to underreport their symptoms while others may face lack of adequate healthcare causing their disorder to go unnoticed [17]. Current diagnostic methodologies can take several appointments for a conclusive diagnosis, which can be time-consuming and cost-prohibitive. One study discovered that white children were twice as likely than black children to get diagnosed with ADHD, boys were five times more likely to be evaluated for ADHD than girls, and poverty was the most pervasive barrier in ADHD diagnosis and care [18]. The same study highlighted that although 88% of the participants were recognized as having a problem, only 32% received an ADHD diagnosis. Therefore, alternative methods of diagnoses that are more quantitative in nature may be more effective, accurate, and cost-efficient.

**Current Diagnostic Technologies**
In recent years, technology has been developed for quantitative diagnosis of psychiatric disorders such as Autism Spectrum Disorder (ASD) and ADHD in children. However, while some of them have taken gender into consideration, none of the studies reviewed in this paper appear to have considered the race of their participants. One study employed the use of virtual reality integrated with multiple sensor technologies to collect physiological data such as eye and head movements to diagnose ADHD in students [19]. It is important to note that this study does not mention the gender and race of the participants, and therefore it is difficult to draw conclusions on its effectiveness in eliminating bias. Other studies took the gender of participants into consideration but not their race and employed various machine learning techniques to identify and classify mental disorders including ADHD. The results of these studies displayed no statistically significant differences in gender [20][21][22]. However, the authors did not address any selection biases that may exist for female participants. Since women and girls with ADHD are more likely to compensate for the debilitating impairments caused by their disorder with prosocial behaviors, researchers must take into consideration that women who have received a professional ADHD diagnosis often have other behavioral and emotional issues which may impact the accuracy of their results [23].

**The Role of HCI Research**
With the exception of the previously highlighted study using virtual classrooms to diagnose ADHD [19], existing HCI literature on ADHD focuses primarily on rehabilitation and assistive technologies for people who have already been diagnosed with the disability [24][25][26]. Other HCI papers have analyzed studies with machine learning models that exacerbate social inequalities, but they have not focused specifically on ADHD diagnosis [27][28]. HCI researchers can contribute to decreasing racial and gender bias in ADHD diagnosis by 1) reproducing the studies mentioned in the previous section while taking into account the race and gender of the participants and how stereotypes and biases may impact their results 2) conducting new studies to analyze the usefulness of quantitative diagnostic technology for women and people of color and identifying their impact on eliminating bias. Such studies will help detect any issues that may exist in the aforementioned papers which can be remedied by computer scientists.

## Conclusions and Suggestions

The gender and racial bias in currently existing technology used to diagnose ADHD still remains to be studied. Studies that report no statistically significant gender bias and high accuracy rates have not analyzed the possibility of racial bias. None of the studies reviewed in this paper took into consideration the impact racial and gender stereotypes might have had on their participant selection and results, or the race of their participants. Identifying the existence of bias in the algorithms and datasets of diagnostic tools used to identify psychiatric disorders such as ADHD is an important topic to combat racial and gender disparities in mental healthcare. HCI researchers can contribute to this field by conducting studies that take gender bias and racial bias into consideration when designing new diagnostic technologies or replicating previous studies. Examples of such biases include misdiagnosing black boys with ODD instead of ADHD due to racial stereotypes and women's tendency to compensate for debilitating disabilities by adapting prosocial behaviors.

## References

[1] Bloom B, Cohen RA, Freeman G. Summary health statistics for U.S. children: National Health Interview Survey, 2009. National Center for Health Statistics. Vital Health Stat 10(247). 2010.

[2] Morgan P, Staff J, Hillemeier M, Farkas G, Maczuga S. Racial and Ethnic Disparities in ADHD Diagnosis From Kindergarten to Eighth Grade Pediatrics Jul 2013, 132 (1) 85-93; DOI: 10.1542/peds.2012-2390

[3] Ballentine, K. L. (2019). Understanding Racial Differences in Diagnosing ODD Versus ADHD Using Critical Race Theory. *Families in Society*, *100*(3), 282–292.

[4] Foley D. ADHD and kids: The truth about attention deficit hyperactivity disor-der. Time 2018. http://time.com/growing-up-with-adhd/

[5] J Kendall, & D Hatton (2003). Racism as a Source of Health Disparity in Families with Children with Attention Deficit Hyperactivity Disorder. ANS. Advances in nursing science. 25. 22-39. 10.1097/00012272-200212000-00003.

[6] Moody, M. From Under-Diagnoses to Over-Representation: Black Children, ADHD, and the School-To-Prison Pipeline. J Afr Am St 20, 152–163 (2016). https://doi.org/10.1007/s12111-016-9325-5

[7] Belcher, J. R. (2014). Attention Deficit Hyperactivity Disorder in Offenders and the Need for Early Intervention. *International Journal of Offender Therapy and Comparative Criminology*, *58*(1), 27–40. https://doi.org/10.1177/0306624X12465583

[8] Ingram, S., Hechtman, L., Morgenstern, G. (1999). Outcome issues in ADHD: Adolescent and adult long-term outcome. Mental Retardation and Developmental Disabilities Research Reviews, 5(3), 243-250. doi:10.1002/(SICI)1098-2779(1999)5:3<243::AID-MRDD11>3.0.CO;2-D

[9] Elkins, I. J., Saunders, G., Malone, S. M., Keyes, M. A., Samek, D. R., McGue, M., & Iacono, W. G. (2018). Increased Risk of Smoking in Female Adolescents Who Had Childhood ADHD. *The*

*American journal of psychiatry*, *175*(1), 63–70. doi:10.1176/appi.ajp.2017.17010009

[10] Wajszilber, D., Santiseban, J. A., & Gruber, R. (2018). Sleep disorders in patients with ADHD: impact and management challenges. *Nature and science of sleep*, *10*, 453–480. doi:10.2147/NSS.S163074

[11] Bagwell, C. L., Molina, B. S. G., Pelham, W. E., Hoza, B. (2001). Attention-deficit hyperactivity disorder and problems in peer relations: Predictions from childhood to adolescence. Journal of the American Academy of Child & Adolescent Psychiatry, 40, 1285-1292. doi:10.1097/00004583-200111000-00008

[12] Murphy, K., Barkley, R. A. (1996). Attention deficit hyperactivity disorder adults: Comorbidities and adaptive impairments. Comprehensive Psychiatry, 37, 393-401.

[13] American Psychiatric Association. (2013). *Diagnostic and statistical manual of mental disorders (DSM-5®)*. American Psychiatric Pub.

[14] Gualtieri, C. T., & Johnson, L. G. (2005). ADHD: Is Objective Diagnosis Possible?. Psychiatry (Edgmont (Pa. : Township)), 2(11), 44–53.

[15] Coker, T. R., Elliott, M. N., Toomey, S. L., Schwebel, D. C., Cuccaro, P., Emery, S. T., ... & Schuster, M. A. (2016). Racial and ethnic disparities in ADHD diagnosis and treatment. Pediatrics, 138(3), e20160407.

[16] Skogli, E.W., Teicher, M.H., Andersen, P.N. *et al.* ADHD in girls and boys – gender differences in co-existing symptoms and executive function measures. *BMC Psychiatry* 13, 298 (2013). https://doi.org/10.1186/1471-244X-13-298

[17] Wood, W. L. M., Lewandowski, L. J., & Lovett, B. J. (2019). Profiles of Diagnosed and Undiagnosed College Students Meeting ADHD Symptom Criteria. *Journal of Attention Disorders*. https://doi.org/10.1177/1087054718824991

[18] Bussing, R., Zima, B.T., Gary, F.A. *et al.* Barriers to detection, help-seeking, and service use for children with ADHD symptoms. *The Journal of Behavioral Health Services & Research* 30, 176–189 (2003). https://doi.org/10.1007/BF02289806

[19] Yeh, Shih-Ching & Tsai, Chia-Fen & Fan, Yao-Chung & Liu, Pin-Chun & Rizzo, Albert. (2012). An innovative ADHD assessment system using virtual reality. 2012 IEEE-EMBS Conference on Biomedical Engineering and Sciences, IECBES 2012. 78-83. 10.1109/IECBES.2012.6498026.

[20] A. Leontyev, T. Yamauchi and M. Razavi, "Machine Learning Stop Signal Test (ML-SST): ML-based Mouse Tracking Enhances Adult ADHD Diagnosis," *2019 8th International Conference on Affective Computing and Intelligent Interaction Workshops and Demos (ACIIW)*, Cambridge, United Kingdom, 2019, pp. 1-5. doi: 10.1109/ACIIW.2019.8925073

[21] Sutoko S, Monden Y, Tokuda T, et al. Distinct Methylphenidate-Evoked Response Measured Using Functional Near-Infrared Spectroscopy During Go/No-Go Task as a Supporting Differential Diagnostic Tool Between Attention-Deficit/Hyperactivity Disorder and Autism Spectrum Disorder Comorbid Children. Front Hum Neurosci. 2019;13:7. Published 2019 Feb 8. doi:10.3389/fnhum.2019.00007

[22] Kwan, C. J., & Pil, J. S. (2018). Clinical Application Of Vibraimage Technology And System For Screening Of Adhd Children. *ELSYS Corp., European Academy of Natural Sciences (EANS), Russian Biometric Association (RBA), AI Burnazyan Federal Medical and Bio*, 178.

[23] Mowlem, F., Agnew-Blais, J., Taylor, E., & Asherson, P. (2019). Do different factors influence whether girls versus boys meet ADHD diagnostic criteria? Sex differences among children with high ADHD symptoms. *Psychiatry research*, *272*, 765–773. doi:10.1016/j.psychres.2018.12.128

[24] J. E. Muñoz, D. S. Lopez, J. F. Lopez and A. Lopez, "Design and creation of a BCI videogame to train sustained attention in children with ADHD," *2015 10th Computing Colombian Conference (10CCC)*, Bogota, 2015, pp. 194-199. doi: 10.1109/ColumbianCC.2015.7333431

[25] Lin, Chien-Yu & Yu, Wen-Jeng & Chen, Wei-Jie & Huang, Chun-Wei & Lin, Chien-Chi. (2016). The Effect of Literacy Learning via Mobile Augmented Reality for the Students with ADHD and Reading Disabilities. 9739. 103-111. 10.1007/978-3-319-40238-3_11.

[26] Smit, Dorothé & Bakker, Saskia. (2015). BlurtLine: A Design Exploration to Support Children with ADHD in Classrooms. 10.1007/978-3-319-22723-8_37.

[27] Allison Woodruff, Sarah E Fox, Steven Rousso-Schindler, and Jeffrey Warshaw. 2018. A qualitative exploration of perceptions of algorithmic fairness. In Proceedings of the 2018 CHI Conference on Human Factors in Computing Systems (CHI 2018). ACM, 656

[28] Aaron Springer, J. Garcia-Gathright, and Henriette Cramer. 2018. Assessing and addressing algorithmic bias—But before we get there. In Proceedings of the AAAI 2018 Spring Symposium: Designing the User Experience of Artificial Intelligence.